\documentclass[iop]{emulateapj}
\usepackage{graphics,graphicx}
\newcommand\ebv{E(\bv)}
\newcommand\msun{\ensuremath{M_\sun}}

\slugcomment{Astronomical Journal MS \#340334 Version 2} 
\shorttitle{Nova-Like Variable in the Kepler Field}
\shortauthors{Williams et al.}

\begin{document}
\title{Discovery of a Nova-Like Cataclysmic Variable in the Kepler Mission Field}

\author{Kurtis A. Williams\footnote{NSF Astronomy \& Astrophysics
Postdoctoral Fellow}}
\affil{Department of Astronomy, University of Texas, 1 University Station C1400,
 Austin, TX 78712, USA}
\email{kurtis@astro.as.utexas.edu}

\author{Domitilla de Martino}
\affil{INAF--Osservatorio di Capodimonte, Moiarello 16, 80131, Napoli, Italy}

\author{Roberto Silvotti}
\affil{INAF--Osservatorio Astronomico di Torino, strada dell'Osservatorio 20, 
10025 Pino Torinese, Italy}

\author{Ivan Bruni}
\affil{INAF--Osservatorio Astronomico di Bologna, via Ranzani 1, 40127 Bologna, 
Italy}

\author{Patrick Dufour}
\affil{D\'epartement de physique, Universit\'e de Montr\'eal \\ Montr\'eal, 
QC H3C 3J7, Canada}

\author{Thomas S.~Riecken, Martin Kronberg, Anjum Mukadam}
\affil{Astronomy Department, Box 351580, University of Washington, Seattle, 
WA 98115, USA}

\and

\author{G.\ Handler}
\affil{Institut f\"ur Astronomie, Universit\"at Wien, T\"urkenschanzstrasse 17, 
A-1180 Wien, Austria}

\begin{abstract}
  We announce the identification of a new cataclysmic variable star in
  the field of the Kepler Mission, KIC J192410.81$+$445934.9.  This
  system was identified during a search for compact pulsators in the
  Kepler field.  High-speed photometry reveals coherent
  large-amplitude variability with a period of 2.94 h.  Rapid,
  large-amplitude quasi-periodic variations are also detected on time
  scales of $\approx 1200$ s and $\approx 650$ s.  Time-resolved
  spectroscopy covering one half photometric period shows shallow,
  broad Balmer and \ion{He}{1} absorption lines with bright emission
  cores as well as strong \ion{He}{2} and Bowen blend emission.
  Radial velocity variations are also observed in the Balmer and
  \ion{He}{1} emission lines that are consistent with the photometric
  period.  We therefore conclude that KIC J192410.81$+$445934.9 is a
  nova-like variable of the UX UMa class in or near the period gap,
  and it may belong to the rapidly growing subclass of SW Sex systems.
  Based on 2MASS photometry and companion star models, we place a
  lower limit on the distance to the system of $\sim 500$ pc.  Due to
  limitations of our discovery data, additional observations including
  spectroscopy and polarimetry are needed to confirm the nature of
  this object. Such data will help to further understanding of the
  behavior of nova-like variables in the critical period range of
  $3-4$ h, where standard cataclysmic variable evolutionary theory
  finds major problems.  The presence of this system in the Kepler
  mission field-of-view also presents a unique opportunity to obtain a
  continuous photometric data stream of unparalleled length and
  precision on a cataclysmic variable system.
\end{abstract}
\keywords{binaries: close --- novae, cataclysmic variables}

\section{Introduction\label{sec.intro}}
Cataclysmic variables (CVs) are close binary systems in which a white
dwarf (WD) accretes material from a companion star.  The secondary
star is a late-type dwarf filling its Roche lobe, and, in most cases,
the overflowing material forms an accretion disk around the WD
primary.

One class of CVs is the nova-like (NL) system.  Historically, a NL
system not showing magnetic characteristics is classified as a
\object{UX UMa} star \citep{1996ASSL..208....3D}.  Non-magnetic NLs
exhibit the same variety of optical spectra as the dwarf novae in
outburst, ranging from pure emission-line spectra (high inclination)
to pure absorption-line spectra (low inclination).  Therefore, NLs are
characterized by relatively high mass accretion rates.  

\begin{deluxetable}{cccl}
\tablecolumns{4}
\tablewidth{0pt}
\tablecaption{Astrometric and Photometric Properties of
  KIC J1924$+$4459\label{tab.kic_data}}
\tablehead{\colhead{Parameter} & \colhead{Value} & \colhead{Date} & \colhead{Source}} 
\startdata
RA(J2000)      & $19^{\rm h}24^{\rm m} 10\fs 82$ & \nodata & KIC10\tablenotemark{a}\\
Dec(J2000)     & $+44\degr 59\arcmin 34\farcs 9$ & \nodata & KIC10\tablenotemark{a} \\
$\mu_{\rm RA}$  & $24\pm 8$ mas yr$^{-1}$  & 1980.90 & USNO-B1.0 \\
$\mu_{\rm Dec}$ & $34\pm 17$ mas yr$^{-1}$  & 1980.90 & USNO-B1.0 \\ 
$B_p$          & 15.22 & 1951.5195 & USNO-B1.0 \\
$R_p$          & \nodata & 1951.5195 & USNO-B1.0 \\
$B_p$          & 15.59 & 1976.56 & YB6\tablenotemark{b} \\
$V_p$          & 15.79 & 1976.56 & YB6\tablenotemark{b} \\
$B_p$          & 15.30 & 1988.4695 & USNO-B1.0 \\
$R_p$          & 15.36 & 1991.5795 & USNO-B1.0 \\
$I_p$          & 15.72 & 1990.4463 & USNO-B1.0 \\
$J$            & $16.839\pm 0.137$ & 1998.436 & 2MASS \\
$H$            & $\geq 16.736$ & 1998.436 & 2MASS \\
$K$            & $\geq 16.536$ & 1998.436 & 2MASS \\
$g$            & 16.182  & 2007.1\tablenotemark{c} & KIC10\tablenotemark{a} \\
$g-r$          & $-0.116$  & 2007.1\tablenotemark{c} & KIC10\tablenotemark{a} \\
$r-i$          & $-0.091$  & 2007.1\tablenotemark{c} & KIC10\tablenotemark{a} \\
$r-z$          & $-0.214$  & 2007.1\tablenotemark{c} & KIC10\tablenotemark{a} \\
\enddata
\tablenotetext{a}{Kepler Input Catalog v.~10, \citealt{2008PhST..130a4034L}}
\tablenotetext{b}{USNO Yellow-Blue Catalog v.~6, D.~Monet (private communications)}
\tablenotetext{c}{KIC10 photometry is a weighted average of data obtained 2006.281, 2007.412, and 2007.486}
\end{deluxetable} 

The classification scheme may improve with consideration of more
precise photometric and spectroscopic characteristics.  Systems in the
3--4 h orbital period range can be identified as \object{VY Scl}
stars when they are found to undergo low states or can be identified
as \object{SW Sex} stars when their spectra display single-peaked
emission features irrespective of inclination angle with absorption
components variable along the orbital period
\citep{Thorstensen91,1995cvs..book.....W,1996ASSL..208....3D}.
Recently, the SW Sex class has been claimed to be the dominant
population of the 3--4 h orbital period CV systems
\citep[e.g.,][]{2007MNRAS.377.1747R,2007MNRAS.374.1359R}.  NLs,
generally not (yet) known to display the above characteristics are
classified as UX UMa stars from the presence of strong absorption
features in Balmer lines with superposed narrow emission cores as well
as emission lines of \ion{He}{1} and \ion{He}{2}.  The detection of
emission in these lines are clear signs of accretion and therefore
cannot be mistaken with non-accreting WD$+$M binaries.

Although the historical definition of NL systems is of a non-magnetic
CV, there is considerable evidence that magnetism is present in at
least some NLs, and that magnetism may be the force behind many of the
defining characteristics of SW Sex systems.  Based on profiles of
emission lines during eclipses, \citet{Williams89} proposed that
magnetic accretion columns could be present in NLs.  Magnetism
was also invoked to explain the complex behavior in the SW Sex star
\object{V795 Her} \citep[e.g.,][]{Thorstensen86,Casares96} and
confirmed with the detection of variable circular polarization in the
system \citep{Rodriguez02}.  Variable circular polarization has also
been detected in the SW Sex stars \object{LS Peg} \citep{Rodriguez01}
and \object{RX J1643.7+3402} \citep{Rodriguez09}.  Other evidence of
magnetism has also been observed in multiple SW Sex systems, including
emission line flaring in systems such as \object{BT Mon} \citep{Smith98} and
\object{V533 Her} \citep{RodMar02} and far UV / X-ray signatures of
magnetism in objects like \object{DW UMa} \citep{Hoard03} and LS Peg
\citep{Baskill05}.  Clearly magnetism plays an important role in many,
if not all, SW Sex stars.

We report on the discovery of a NL, UX UMa-type system, KIC
J192410.81$+$445934.9 (Kepler ID 8751494, hereafter KIC J1924$+$4459), in
the field-of-view of the Kepler Mission.  This discovery was made as
part of an as-yet unsuccessful search for compact pulsators (WDs and
subdwarf B stars) in the Kepler Input Catalog.  Time-series
photometric observations of compact pulsator candidates were obtained
at the TNG, McDonald Observatory, and the BOAO Observatory (Seung
Lee-Kim, private communication) without any clear detection of
pulsations in any of the observed targets.  Some preliminary TNG
results are presented by \citet{2009CoAst.159...97S}.

The detection of a photometric periodicity at 2.94 h in KIC J1924$+$4459,
potentially also detected in the radial velocities of the emission
lines, and the detection of rapid non-periodic variability hints
that this star may belong to the growing class of SW Sex systems.

\section{Photometric Observations and Analysis \label{sec.phot}}
\subsection{Selection and Time-Series Photometry}
\begin{figure}
\begin{center}
\includegraphics[scale=0.4]{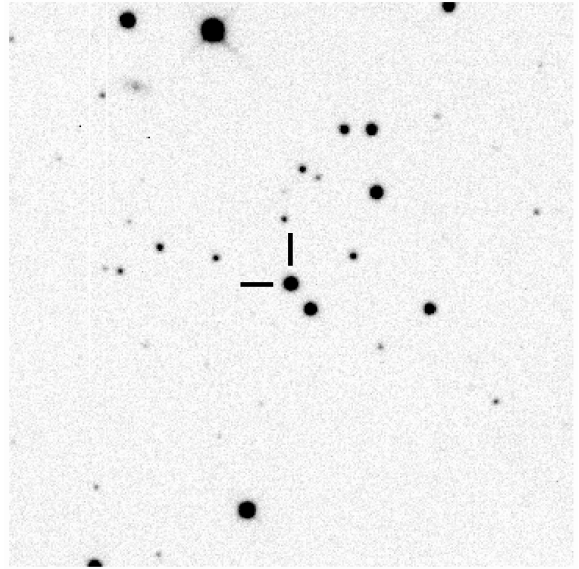}
\end{center}
\caption{Finder chart for KIC J1924$+$4459.  The chart is 2\arcmin ~on
  a side; north is up, east to the left. The system is centered and
  marked with tick marks.  This image was obtained at
  McDonald Observatory through a BG40 filter. \label{fig.finder}}
\end{figure}

Potential pulsating white dwarfs were selected via their location in
the reduced proper motion diagram based on photometric and astrometric
data in the Kepler Input Catalog \citep{2008PhST..130a4034L}.  In
particular, KIC J1924$+$4459 was flagged as a potential ZZ Ceti pulsator; its
astrometric and photometric properties from the Kepler Input Catalog
are given in Table \ref{tab.kic_data}; a finder chart is given in
Figure \ref{fig.finder}.

Initial time-series observations were obtained 24 August 2008 on the
3.6-m TNG telescope with the DOLORES imaging spectrograph. The imaging
data were taken with the $g$-band filter with exposure times of 8
s; a short 0.98 h run revealed flux variations of $\gtrsim
10\%$.  Photometric data were reduced using standard procedures in
IRAF\footnote{IRAF is distributed by the National Optical Astronomy
  Observatories, which are operated by the Association of Universities
  for Research in Astronomy, Inc., under cooperative agreement with
  the National Science Foundation}, including bias, flat-fielding, and
sky subtraction.  Aperture photometry was obtained for the target and
several comparison stars.  The flux ratios were obtained by dividing
the counts of the target by the best combination (weighted mean) of
three reference stars.  These ratios were converted to fractional
intensities by dividing the mean flux ratio.  Barycentric corrections
were also applied.

Additional time-series data were obtained at McDonald Observatory and
the Bologna Astronomical Observatory.  The McDonald observations were
obtained with the Argos high-speed photometer on the 2.1-meter Struve
Telescope; Argos uses a prime-focus, back-illuminated frame-transfer
CCD to obtain uninterrupted time-series photometry
\citep{2004ApJ...605..846N}. Data were obtained through a BG40 filter
with individual exposure times of 10 seconds; runs of 2.57 h and 2.06
h were obtained on UT 2008 September 5 and 6, respectively.  Data
were reduced in the manner described by \citet{2005ApJ...625..966M}
and \citet{2008ApJ...676..573M} using the IRAF package \emph{ccd\_hsp}
\citep{Odonoghue00}.

The Loiano/Bologna observations were obtained on UT 2008 September 5
and 7 at 1.5-meter Loiano telescope equipped with a BFOSC detector
operated without a filter.  Integration times were 40 s on September
5 and 45 s on September 7.  A total of 332 and 269 frames were
acquired during the 5.3 h and 4.7 h runs of the two nights,
respectively. The data were reduced in the same manner as the TNG
data.

\begin{figure}
\begin{center}
\includegraphics[angle=270, width=0.49\textwidth]{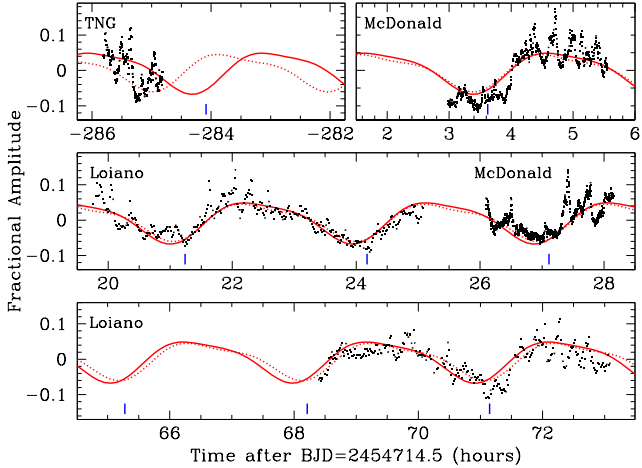}
\end{center}
\caption{Time series photometry for KIC J1924$+$4459 from UT 24 August
  2008 through UT 7 September 2008.  Labels indicate the source of the
  data; the heavy curve is the best-fitting sinusoid plus first
  harmonic (see Section \ref{sec.discuss.phot}) from all data; the
  dotted curve is the best-fitting sinusoid plus first harmonic from
  the Loiano data alone.  Time is given in hours relative to BJD
  2454714.5; the horizontal scale in each panel is the
  same. Vertical tick marks indicate the predicted times of potential
  grazing eclipses in the Loiano data based on a potential grazing
  eclipse at $t=24.175$ h.  \label{fig.ts_phot}}
\end{figure}

The time-series data are shown in Figure \ref{fig.ts_phot}. 

\subsection{Photometric variability analysis\label{sec.discuss.phot}}
\subsubsection{Short-term variability}
The light curves display large amplitude variations of 0.12 mag that
appear to be periodic.  The light curve shape is not sinusoidal, but
shows sharp rises and structured decays.  We then performed a Scargle
analysis of the Loiano data that revealed a strong peak at 94.4
$\mu$Hz and a secondary peak at 188 $\mu$Hz.  The latter peak is close
to twice the frequency of the strong peak.  We therefore fitted a
composite sinusoidal function with a fundamental period of $2.9422\pm
0.0016$ h (frequency $\nu =8.1560\pm 0.0051\, {\mathrm d}^{-1}$ ) and the
first harmonic.

\begin{figure}
\begin{center}
\includegraphics[angle=270, width=0.45\textwidth]{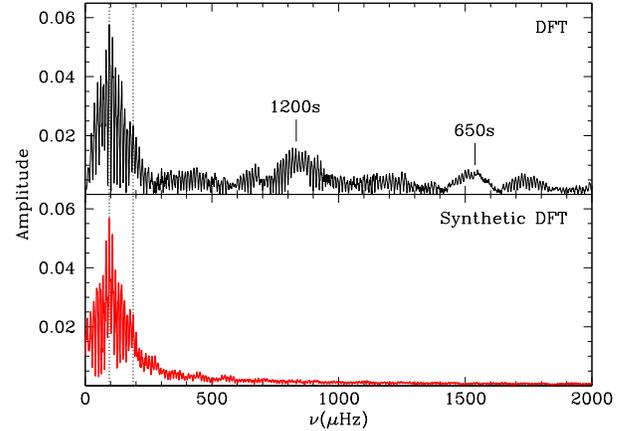}
\end{center}
\caption{Discrete Fourier transforms of the combined McDonald and
  Loiano data sets (top) and of the synthetic light curve generated
  using the sinusoidal fits including the fundamental and first
  harmonic to the combined McDonald and Loiano photometry (bottom).
  Vertical lines indicate the frequencies of the fundamental and first
  harmonic.  Also indicated in the DFT are two broad peaks of
  power likely related to the quasi-coherent variations observed
  in the light curves.  \label{fig.dft}}
\end{figure}

As the amplitudes of the variability were similar in both the
unfiltered Loiano photometry and the BG40 McDonald photometry, and as
the two data sets were interleaved, we combined the two data sets to
more precisely constrain the periodicities in the data.  Due to the
large gap between the discovery photometry and our subsequent
follow-up and the short duration of observation, we did not include
the discovery TNG data in the combined analysis.  A discrete Fourier
transform (DFT) of the combined data (Fig.~\ref{fig.dft}) shows a strong
peak very near the frequency of the Loiano data peak.  Again, we
simultaneously fit a fundamental frequency and its first harmonic
\begin{equation}
f=A_0\sin(2\pi \frac{t-t_{0,\rm phot}}{P_{\rm phot}}) + A_1\sin(4\pi
\frac{t-t_1}{P_{\rm phot}})\, ,
\label{eqn.phot}
\end{equation}
where $f$ is the fractional amplitude. For the fundamental frequency,
we derive a photometric period $P_{\rm phot}=2.9358\pm 0.0017$ h
($\nu= 8.1749\pm 0.0070 \,{\mathrm d}^{-1}$) with an
amplitude $A_0=0.0556\pm 0.0014$ and $t_{0,\rm phot} ({\rm BJD}) = 
2454715.52547(74)$.  For the first harmonic, $A_1=0.0133\pm 0.0014$
and $t_1 ({\rm BJD}) = 2454715.57688(148)$. 
Figure \ref{fig.ts_phot} shows this fit overplotted on the
photometric data.

Although the formal errors indicate a significant difference between
this period and that derived from the Loiano photometry alone, we
believe the true error in the period is larger than the formal errors.
In Figure \ref{fig.ts_phot}, the model light curves for both periods
are plotted over the observed photometry.  Even when considering the
TNG data, neither period is obviously preferred.  In addition, light
curves folded at both periods look qualitatively similar.  We
therefore consider a realistic estimate on the systematic error in the
period to be 0.0064 h, the difference between the two photometric
periods.  Additional observations will be needed to determine the
photometric period more precisely. If the photometric period of 2.94 h
is the orbital period of the system, then KIC J1924$+$4459 lies within
the $2-3$ h CV period gap. 

No deep eclipses are observed.  Small dips which could be grazing
eclipses are visible just after minimum light in the Loiano
data. These have been indicated in Figure \ref{fig.ts_phot} with
vertical tick marks.  The location of the tick mark at T=24.175 h
after BJD 2454714.5 was set by visual inspection; all other tick marks
are spaced by integer multiples of the photometric period of 2.9358 h.
Thse dips are consistent with the photometric period throughout both Loiano
data sets, but are not seen in the McDonald data.  A longer data set, such
as that being obtained by the Kepler mission, therefore will be
required to address this issue.

After removing this long-period variability, the residual light curves
from all four nights of data also show significant amplitude ($\sim
5\%$) variability on time scales of $\sim 650s$ and $\sim 1200s$.
Significant power is visible at these frequencies in the DFTs of the
combined data (Fig.~\ref{fig.dft}) and of each individual run.
However, this variability is not fully coherent, even within a single
observation, indicating that these are not non-radial pulsations on
the surface of the WD primary.  

\subsubsection{Long-Term Variability}
We have been able to identify KIC J1924$+$4459 in several archival
catalogs, including both epochs of the Palomar Observatory Sky Survey
(POSS), the Lick Northern Proper Motion study, and 2MASS.  Historic
photometric data culled from the the Naval Observatory Merged
Astrometric Dataset (NOMAD) are presented in Table \ref{tab.kic_data}.
We note that the USNO-A2 catalog does not resolve KIC J1924$+$4459 and
a similarly-bright optical companion star; we therefore do not include
those data in this summary.  Visual inspection of both epochs of the
POSS reveals significant proper motion of KIC J1924$+$4459 relative to
the optical companion, so these two sources are likely unrelated.


Even allowing for uncertainty in the photographic plate magnitudes,
residual contamination from the neighbor star, and the uncertain phase
of the recent digital photometry, it appears that KIC J1924$+$4459 may
have faded by $\sim 0.5-1$ magnitude between the second-epoch POSS
observations of the early 1990s and the 2MASS/KIC observations of the
late 1990s and mid-2000s.  Such fading episodes are observed in NLs of
the \objectname{VY Scl} subclass.  Given the uncertainties, however,
we cannot conclusively state whether KIC J1924$+$4459 entered a VY
Scl-like low state in the mid- to late 1990s.


\section{Spectroscopic Analysis}
\subsection{Spectroscopic Observations}
Spectroscopic observations of KIC J1924$+$4459 were obtained with the
Blue Channel spectrograph on the MMT on UT 2008 September 24 and 25.
On the first night, a single 300 s exposure revealed broad hydrogen
absorption with narrow emission cores; this led us to an initial
interpretation that this object could be a WD$+$M binary.

On the second night, significant additional spectroscopy was obtained
when high winds precluded observations of the spectroscopic run's
primary targets. Data were obtained with a 500 grooves per millimeter
grating blazed at 5410\AA\ and a 1\arcsec-wide slit; no order blocking
filter was used. Wavelength coverage extends from 3200\AA\ to 6300\AA.
\ The resulting spectral resolution, measured from isolated night-sky
emission lines, gives a Gaussian full-width at half maximum (FWHM) of
3.8\AA.

\begin{figure}
\begin{center}
\includegraphics[angle=270,bb=275 1 597 785,width=0.45\textwidth]{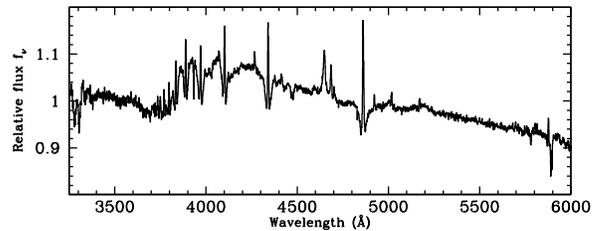}
\end{center}
\caption{Averaged optical spectrum of KIC J1924$+$4459. The shallow,
  broad Balmer and \ion{He}{1} absorption lines with emission cores,
  the strong \ion{He}{2} and Bowen blend emission lines, and the
  appearance of the Balmer jump in emission are all typical of
  nova-like variables of the UX UMa class. \label{fig.spec}}
\end{figure}

\tabletypesize{\scriptsize}
\begin{deluxetable*}{ccccccccccc}
\tablecolumns{11}
\tablewidth{0pt}
\tablecaption{Equivalent Widths of Prominent Lines in KIC J1924$+$4459\label{tab.ews}}
\tablehead{\colhead{Obs.\ Midpt.} & \colhead{S/N} & 
\multicolumn{2}{c}{H$\beta$} & \multicolumn{2}{c}{H$\gamma$} & 
\multicolumn{2}{c}{\ion{He}{1} $\lambda 4471$} & 
\colhead{\ion{He}{1}} & 
\colhead{\ion{He}{2}} & \colhead{\ion{N}{3} / \ion{C}{3}} \\
 (BJD$-$& & total & emiss. & total & emiss. & total & emiss.&  $\lambda 5876$ &  $\lambda 4686$ & \\
  $2450000)$  & & (\AA) & (\AA) & (\AA) & (\AA) & (\AA) & (\AA) & (\AA) & (\AA) & (\AA) }
\startdata
4734.60556 & 93 & $1.49\pm 0.29$ & $-2.48\pm 0.04$ & $1.65\pm 0.10$ & $-1.38\pm 0.02$ & $0.31\pm 0.03$ & $-0.17\pm 0.01$ & $-0.45\pm 0.03$ & $-1.08\pm 0.18$ & $-1.23\pm 0.07$ \\
4734.61133 & 90 & $1.01\pm 0.27$ & $-0.74\pm 0.09$ & $1.41\pm 0.32$ & $-1.31\pm 0.02$ & $0.40\pm 0.05$ & $-0.14\pm 0.01$ & $-0.38\pm 0.03$ & $-0.74\pm 0.09$ & $-1.30\pm 0.10$ \\
4734.61713 & 95 & $1.59\pm 0.25$ & $-1.92\pm 0.05$ & $1.24\pm 0.22$ & $-1.42\pm 0.01$ & $0.25\pm 0.05$ & $-0.11\pm 0.02$ & $-0.38\pm 0.03$ & $-0.42\pm 0.09$ & $-1.40\pm 0.13$ \\
4734.62292 &111 & $1.18\pm 0.21$ & $-2.23\pm 0.05$ & $1.12\pm 0.17$ & $-1.30\pm 0.01$ & $0.46\pm 0.07$ & $-0.11\pm 0.02$ & $-0.23\pm 0.06$ & $-0.41\pm 0.09$ & $-1.38\pm 0.10$ \\
4734.62877 &104 & $1.31\pm 0.07$ & $-1.84\pm 0.04$ & $0.92\pm 0.13$ & $-1.42\pm 0.03$ & $0.75\pm 0.09$ & $-0.07\pm 0.01$ & $-0.36\pm 0.08$ & $-0.50\pm 0.08$ & $-1.18\pm 0.17$ \\
4734.63649 &105 & $0.95\pm 0.16$ & $-1.89\pm 0.09$ & $1.15\pm 0.05$ & $-1.27\pm 0.03$ & $0.49\pm 0.09$ & $-0.03\pm 0.02$ & $-0.08\pm 0.15$ & $-0.57\pm 0.09$ & $-1.17\pm 0.07$ \\
4734.64229 &101 & $1.46\pm 0.16$ & $-1.86\pm 0.07$ & $1.34\pm 0.20$ & $-1.04\pm 0.04$ & $0.56\pm 0.04$ & $-0.02\pm 0.03$ & $-0.07\pm 0.14$ & $-0.45\pm 0.07$ & $-1.01\pm 0.13$ \\
4734.64807 & 99 & $1.14\pm 0.27$ &$-1.86\pm 0.04$&$1.47\pm 0.11$&$-1.02\pm 0.05$&$0.51\pm 0.05$&$\phantom{-}0.00\pm 0.04$& $-0.02\pm 0.14$ & $-0.23\pm 0.05$ & $-1.30\pm 0.11$ \\
4734.65769 & 94 & $1.18\pm 0.27$ & $-2.05\pm 0.09$ & $1.45\pm 0.19$ & $-1.25\pm 0.02$ & $0.51\pm 0.09$ & $-0.03\pm 0.02$ & $-0.11\pm 0.14$ & $-0.41\pm 0.04$ & $-1.55\pm 0.12$ \\
Averaged   &211 & $1.45\pm 0.10$ & $-1.97\pm 0.03$ & $1.26\pm 0.20$ & $-1.19\pm 0.02$ & $0.50\pm 0.05$ & $-0.05\pm 0.03$ & $-0.25\pm 0.09$ & $-0.48\pm 0.04$ & $-1.29\pm 0.04$ \\
\enddata
\end{deluxetable*}
\tabletypesize{\footnotesize}

Nine 480-s exposures were obtained over 1.4 hours before winds
forced dome closure, corresponding to nearly half of the suspected
orbital period. In Figure \ref{fig.spec}, we
show the average spectrum from these nine exposures. The spectrum is
dominated by broad Balmer absorption with emission lines at the
line core.  \ion{He}{1} $\lambda 4471$ is also observed as a broad
absorption line with an emission core.  Numerous other weak
\ion{He}{1} lines are observed in emission only.  Strong emission
lines of \ion{He}{2} $\lambda 4686$ and the \ion{C}{3}/\ion{N}{3}
Bowen blend emission near 4648\AA\  are also present.  Equivalent widths
(EWs) of
prominent absorption and emission lines are given in Table
\ref{tab.ews}.

Strong, narrow \ion{Ca}{2} K and \ion{Na}{1} D lines are observed in
absorption. As the source is located near the galactic plane
($b=+13\fdg 46$), these absorption features seem likely to be
interstellar in origin.  Using the dust maps of
\citet{1998ApJ...500..525S}, the estimated color excess due to dust
extinction is $\ebv=0.144$ mag, assuming all extinction is foreground
to the system.  Along the line of sight toward the CV, the total
hydrogen column density is $1.21\times 10^{21}\,{\rm cm}^{-2}$
\citep{1990ARA&A..28..215D}, implying an upper limit $\ebv\leq 0.18$.
The EW of the Na D lines measured from our spectroscopy is $0.57\pm
0.02$\AA; based on the empirical relation of interstellar Na D
strength to $\ebv$ ~of \citet{1997A&A...318..269M}, our measurement
implies a significantly higher $\ebv\approx 0.45\pm 0.15$.  This
suggests that KIC J1924$+$4459 is either reddened substantially by
patchy interstellar extinction, or that significant Na D absorption is
intrinsic to the system.

\subsection{Radial velocity variations}

\begin{figure}
\begin{center}
\includegraphics[angle=270,width=0.49\textwidth]{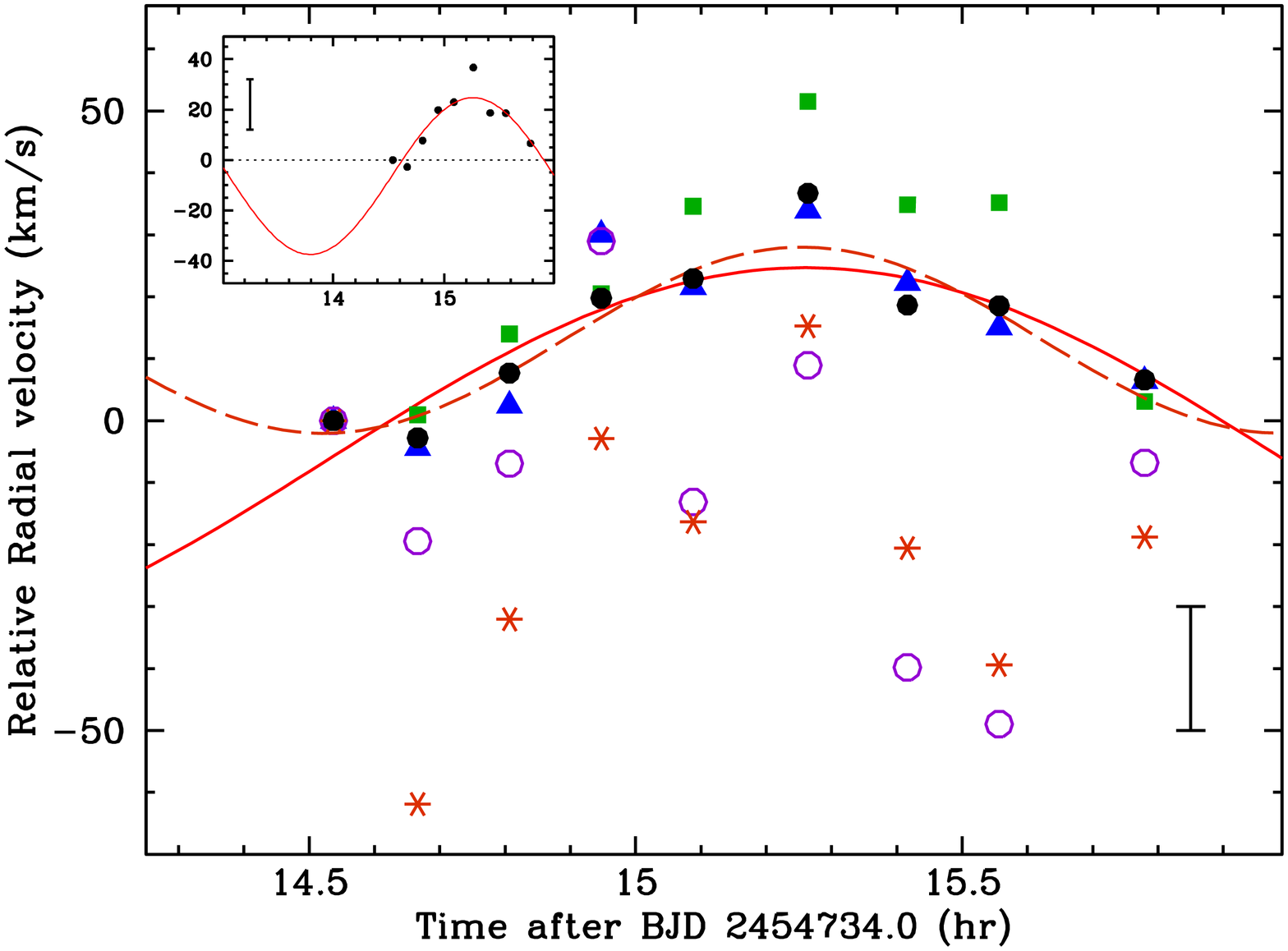}
\end{center}
\caption{Radial velocity measurements for KIC J1924$+$4459 based on
  MMT low-resolution spectroscopy.  Points indicate relative
  velocities for all emission lines (filled circles), H$\beta$
  emission lines only (filled squares), Balmer emission lines of
  H$\delta$ through and including H8 (filled triangles), \ion{He}{2}
  emission (open circles), and the \ion{C}{3}/\ion{N}{3} Bowen blend
  emission (asterisks). The error bar in the lower right shows a
  typical uncertainty ($\pm 10$ km s$^{-1}$) in the relative velocity
  measurements; individual errors are given in Table \ref{tab.rvs}.
  The solid curve is the best-fit sinusoid to all 
  emission lines assuming the orbital period to be equal to the
  photometric period; the dashed curve is the best fit sinusoid to all
  data assuming an orbital period of half the photometric period.  The
  inset shows one complete photometric cycle; our measurements span
  roughly half the photometric period.  The Balmer emission line
  motion may be indicative of orbital motion.
  \label{fig.rvcurve}}
\end{figure}

We searched for evidence of radial velocity variations by
cross-correlating each individual spectrum with the first spectrum.
Continuum, including the broad Balmer absorption, was fit and
subtracted, so that the cross-correlation results in the emission line
velocity curve.  These relative velocities are listed in Table
\ref{tab.rvs}, and the resulting radial velocity curve is shown in Figure
\ref{fig.rvcurve}. 

\begin{deluxetable*}{cccccc}
\tablewidth{0pt}
\tablecolumns{6}
\tablecaption{Emission line relative radial velocity measurements for KIC J1924$+$4459 \label{tab.rvs}}
\tablehead{\colhead{Obs. Midpt.} & \colhead{All Lines} & \colhead{H$\beta$} & 
  \colhead{H$\delta$--H8} & \colhead{\ion{He}{2} $\lambda 4686$} & \colhead{\ion{N}{3}/\ion{C}{3}} \\
  \colhead{(BJD-2450000)} & & & & &}
\startdata

4734.60556 & \nodata & \nodata & \nodata & \nodata & \nodata \\ 
4734.61133 & $ -2.8\pm 10.3$ & $  1.0\pm 6.5$ & $ -4.4\pm  6.3$ & $ -19.4\pm 11.5$ & $ -61.9\pm 11.1$ \\
4734.61713 & $  7.7\pm  9.4$ & $ 14.0\pm 6.6$ & $  2.4\pm  6.5$ & $  -6.9\pm  7.0$ & $ -32.1\pm 11.4$ \\
4734.62292 & $ 19.8\pm 14.3$ & $ 20.5\pm 4.7$ & $ 30.1\pm  8.4$ & $  29.0\pm 14.5$ & $  -2.9\pm 12.1$ \\
4734.62877 & $ 23.0\pm 10.0$ & $ 34.6\pm 5.9$ & $ 21.6\pm  8.6$ & $ -13.1\pm 13.6$ & $ -16.4\pm 11.0$ \\
4734.63649 & $ 36.7\pm 14.3$ & $ 51.5\pm 5.4$ & $ 33.9\pm 10.1$ & $   9.0\pm 15.5$ & $  15.3\pm  7.4$ \\
4734.64229 & $ 18.7\pm 10.1$ & $ 34.9\pm 5.6$ & $ 22.2\pm 10.6$ & $ -39.8\pm 13.3$ & $ -20.5\pm  9.2$ \\
4734.64807 & $ 18.6\pm 12.7$ & $ 35.2\pm 5.5$ & $ 15.1\pm  7.4$ & $ -49.0\pm 15.6$ & $ -39.4\pm 11.4$ \\
4734.65769 & $  6.6\pm 15.2$ & $  3.1\pm 4.4$ & $  6.4\pm  6.7$ & $  -6.8\pm 21.6$ & $ -18.8\pm  8.2$ \\
\enddata
\tablecomments{Velocities are in km s$^{-1}$ and relative to first observation.}
\end{deluxetable*}

We detect small but significant radial velocity variations. Working
under the assumption that these variations are intrinsic orbital
motions, that the orbital period is equal to the primary photometric
period of $P_{\rm phot}=2.9358$ h, and that the orbit is circular, we
use least-squares fitting to determine an orbital solution of the
form:
\begin{equation}
v_{r,rel} = \gamma + K\cos(2\pi\frac{t-t_{0,\rm spec}}{P})\, .
\end{equation}
The resulting fit (shown in Figure \ref{fig.rvcurve}) gives $K=30.1\pm
11.8$ km s$^{-1}$, $t_{0,\rm spec} ({\rm BJD}) = 2454734.63588(383)$,
and $\gamma=-6.4\pm 9.2$ km s$^{-1}$.  The goodness-of-fit is very
high, with $\chi^2/\nu=0.4$.  We note that the photometric period may
not be the orbital period, but the photometric modulation could be due
to some other phenomenon such as a superhump.  To confirm whether the
photometric period traces the true orbital motion or a superhump
period, longer spectroscopic coverage is needed.

 The absolute systemic radial velocity $\gamma_{\rm abs}$ is obtained by
cross-correlating the continuum-subtracted emission line spectrum with a
artificial zero-velocity Balmer emission spectrum; this velocity was
corrected to the local standard of rest using the standard solar
motion.  This exercise gives $\gamma_{\rm abs}=-32.14\pm 9.2\pm 17.7$
km s$^{-1}$, where the first error is the error arising from the
orbital solution and the second error is the uncertainty in the
absolute heliocentric velocity determination.

Defining zero spectroscopic phase as the red-to-blue velocity
crossing, the time of zero phase $T_{0,\rm spec} (\rm BJD)=
2454734.66646(383)$.  Based on Equation \ref{eqn.phot} and including
the accumulated systematic uncertainty in the photometric period
determination, this corresponds to a photometric phase
$(t-t_{0,\rm phot})/P=0.47\pm 0.30$.  The relatively large
accumulated uncertainty in photometric phase therefore precludes
precise comparison of these spectroscopic data with the photometry at
this time.

Many CVs show different $\gamma$ and $K$ values for different emission
lines due to their different origins within the system.  Therefore, we
also calculated the radial velocity curves subsets of the emission
lines, including H$\beta$, the higher-order Balmer lines
(H$\delta$--H8), \ion{He}{2}, and the Bowen blend line. The relative
velocities for these subsets of lines are listed in Table
\ref{tab.rvs} and plotted in Figure \ref{fig.rvcurve}.

Radial velocity curves were determined by fitting these velocities
with a sinusoid of a fixed period of 2.9358 h. From H$\beta$
velocities alone, the resultant $K = 51.1\pm 4.8$ km s$^{-1}$,
significantly higher than the fit using all the lines, though the fit
is significantly worse ($\chi^2/\nu = 4.4$).  Within the errors,
$\gamma$ is unchanged.  The fit to the higher-order Balmer lines are,
within the errors, consistent with the initial fit obtained using all
emission lines.

The individual sinusoidal fits to \ion{He}{2} and the Bowen blend were
significantly different than the initial and Balmer line fits.  For
\ion{He}{2}, $K=33.9\pm 13.8$ km s$^{-1}$ and $\gamma = -34.0\pm 11.8$
km s$^{-1}$ with $\chi^2/\nu = 4.3$.  For the Bowen blend, $K=15.2\pm
9.1$ km s$^{-1}$, $\gamma=-24.4\pm 6.6$ km s$^{-1}$, and $\chi^2/\nu =
9.0$.  Within the errors, these amplitudes are consistent with one
another, yet of lower amplitude than the H$\beta$ variations, as would
be expected since higher-excitation lines generally originate much closer
to the accreting white dwarf.  However, these fits are only slightly
favored over the weighted means (constant velocity solutions) of
$-11.0\pm 3.6$ km s$^{-1}$ (\ion{He}{2}, $\chi^2/\nu=4.7$) and
$-17.8\pm 2.7$ km s$^{-1}$ (\ion{C}{3}/\ion{N}{3}, $\chi^2/\nu=8.1$).
As the radial velocities of these high-excitation lines are observed
to vary rapidly in some other CVs, particularly in some SW Sex
systems, we again caution that additional observations are necessary
and that these sinusoidal fits may not have any physical meaning.

\subsection{Emission Line Structure and Strength}
The emission lines, including the Balmer series lines, are resolved in
the individual spectra (Table \ref{tab.fwhms}).  For comparison
purposes, the measured FWHM of the unresolved
[\ion{O}{1}] $\lambda5577$ night sky line is included in the table.
Mean velocity widths (FWHM) of the Balmer lines are $ 330\pm 20$ km s$^{-1}$
after correcting for the instrumental resolution; \ion{He}{1} lines
are similarly narrow.  The width of \ion{He}{2} $\lambda4686$ may be
somewhat broader; $\approx 375\pm 15$ km s$^{-1}$. In the summed
spectrum, no evidence for double-peaked emission lines is seen.

\begin{deluxetable*}{ccccccccc}
\tabletypesize{\footnotesize}
\tablewidth{0pt}
\tablecolumns{9}
\tablecaption{Emission line widths in  KIC J1924$+$4459\label{tab.fwhms}}
\tablehead{\colhead{Obs.\ Midpt.} & 
\colhead{[\ion{O}{1}]} & 
\multicolumn{2}{c}{H$\beta$} & \multicolumn{2}{c}{H$\gamma$} & 
\multicolumn{2}{c}{\ion{He}{2} $\lambda 4686$} & 
\colhead{\ion{N}{3}/\ion{C}{3}} \\ 
 (BJD-2450000) & \colhead{$\lambda 5577$} & \colhead{FWHM} & \colhead{$\sigma$} & \colhead{FWHM} & 
 \colhead{$\sigma$} & \colhead{FWHM} & \colhead{$\sigma$} & \colhead{FWHM}  \\
 & \colhead{(\AA)} &  \colhead{(\AA)} & \colhead{km s$^{-1}$} & \colhead{(\AA)} & 
   \colhead{km s$^{-1}$} & \colhead{(\AA)} & \colhead{km s$^{-1}$} & 
   \colhead{(\AA)} }
\startdata
4734.60556 & 3.85 & 7.59 & 172 & 6.22 & 144 & 9.75 & 244 & 15.45 \\
4734.61133 & 3.82 & 6.93 & 152 & 5.80 & 129 & 8.87 & 218 & 14.11 \\
4734.61713 & 3.82 & 6.78 & 147 & 7.28 & 182 & 6.41 & 140 & 17.77 \\
4734.62292 & 3.77 & 7.54 & 171 & 6.21 & 144 & 5.93 & 123 & 14.74 \\
4734.62877 & 3.79 & 6.71 & 145 & 6.72 & 163 & 7.24 & 168 & 13.33 \\
4734.63649 & 3.77 & 6.74 & 146 & 6.85 & 167 & 8.09 & 194 & 12.16 \\
4734.64229 & 3.75 & 7.45 & 168 & 6.31 & 148 & 6.63 & 148 & 11.14 \\
4734.64807 & 3.75 & 7.06 & 156 & 5.58 & 120 & 6.24 & 134 & 13.22 \\
4734.65796 & 3.75 & 6.53 & 139 & 5.78 & 128 & 6.87 & 156 & 20.61 \\
Average    & 3.82 & 6.20 & 128 & 7.39 & 186 & 6.94 & 158 & 14.90 \\
\enddata
\tablecomments{Velocity dispersions, but not FWHMs, are corrected for
  average instrumental broadening of FWHM$=3.82$\AA.  The FWHM of the
  telluric [\ion{O}{1}] auroral line is shown to illustrate the instrumental
  broadening and typical scatter in the measurements.  No velocity
  dispersion is calculated for the \ion{N}{3}/\ion{C}{3} Bowen blend.}
\end{deluxetable*}

\begin{figure}
\begin{center}
\includegraphics[angle=270,width=0.5\textwidth,bb=275 14 596 784]{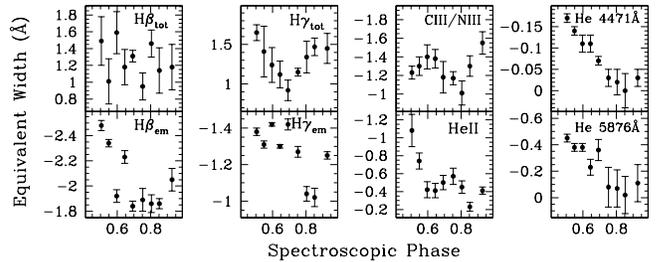}
\end{center}
\caption{Equivalent widths of prominent spectral features as a
  function of spectroscopic phase; zero phase is defined as the
  red-to-blue velocity crossing.  Vertical axes are oriented such that
  stronger measurements, whether of emission or absorption lines, are
  toward the top. Balmer and \ion{He}{1} emission lines show obvious
  trends with phase; the \ion{He}{2} emission line appears quite strong in the
  first two observations. \label{fig.ews}}
\end{figure}

The emission line strengths of all emission lines, with the exception
of the Bowen blend, are observed to vary significantly as a function
of time (Table \ref{tab.ews}, Figure \ref{fig.ews}).  In particular,
the \ion{He}{1} lines ($\lambda 4471$ and $\lambda 5876$) both vanish
toward later times, and the \ion{He}{2} $\lambda 4686$ line is
significantly stronger in the first two exposures than in later
exposures; the H$\beta$ emission line may show similar behavior.  The
Bowen blend emission has a mean EW of $-1.29\pm 0.04$ \AA; all
individual data points are within $2\sigma$ measurement errors of this
mean value, so any intrinsic variability in the EW of this line must
be $\lesssim 0.2$ \AA ~in amplitude.

\begin{figure}
\begin{center}
\includegraphics[width=0.5\textwidth]{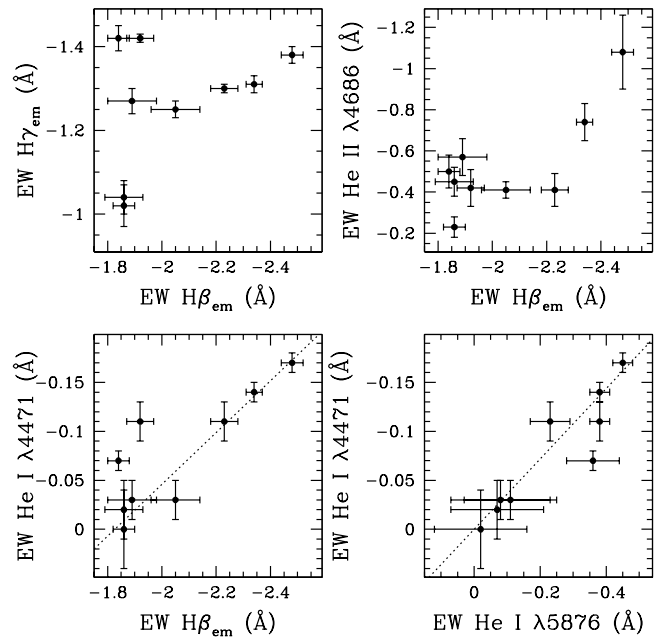}
\end{center}
\caption{Comparison of equivalent widths of various emission lines.
  Dotted lines indicate linear correlations with slopes given in the
  text.  While the correlation of H$\beta$ with H$\gamma$ shows significant
  scatter when H$\beta$ emission is weak, the other correlations are
  strong.  \label{fig.ew_comp}}
\end{figure}

Figure \ref{fig.ew_comp} compares equivalent widths for pairs of
emission lines in four cases where a significant correlation is
observed.  H$\beta$ and H$\gamma$ emission is correlated, as would be
expected, though there is significant scatter at weaker H$\beta$
strengths.  This could be indicative of different emission mechanisms
being observed at different phases, or of systematic problems in
measuring the emission line strengths in one or both lines.  H$\beta$
and \ion{He}{1} $\lambda 4471$ show a nearly linear correlation with
${\rm EW} (\lambda 4471) / \rm EW (H\beta) = 0.26$.
Likewise, the equivalent widths of \ion{He}{1} $\lambda 4471$ and
\ion{He}{1} $\lambda 5876$ are linearly correlated; ${\rm
    EW}(\lambda 4471) / {\rm EW}(\lambda 5876) \approx 0.36$.
Finally, the two spectra with strongest \ion{He}{2} emission also
exhibit the strongest H$\beta$ emission.


Lastly, the strong \ion{He}{2} line might suggest that KIC
J1924$+$4459 could be a source of X-ray emission. Searches of the
available X-ray catalogs via the HEASARC archive fail to result in any
coincident X-ray sources within 10\arcmin ~of the star.

\section{Classification and Implications}
\subsection{A nova-like variable of the UX UMa class}
To summarize our observations, we find that the star KIC J1924$+$4459
is a photometric variable with a period of $\approx 2.94$ h and shows
spectral characteristics of CVs; hence it is a cataclysmic variable in
the $2-3$ h period gap.  Time-series photometry shows no deep
eclipses.  Quasi-periodic variations are also observed on time scales
of $\approx 1200$ s and $\approx 650$ s. The spectrum exhibits Balmer
absorption with resolved, single-peaked emission line cores, as well
as strong \ion{He}{2} and Bowen blend emission. 
Radial velocity variations are observed that 
are consistent with low-amplitude motion and have periods either close to 
the photometric period or close to half of the photometric period.
However, the phase coverage is currently too small to study the
spectral and radial velocity variability in detail.

KIC J1924$+$4459 is clearly a NL variable of the UX UMa class.  The
strong, persistent Balmer absorption features indicate that this is a
system accreting at a high rate. The strength of UX Uma emission features varies
from object to object \citep{1996ASSL..208....3D}. Quasi-periodic
short-term photometric variability is also a common feature observed
in most NLs. KIC J1924$+$4459 is consistent with all of these
descriptors.  We roughly estimate the accretion rate via two methods.
First, the small EW of H$\beta$ emission implies a limit on the disk
luminosity of $M_V\leq 6$, or a lower limit on the accretion rate of
$\dot M\gtrsim 3\times 10^{-10}\,\msun\,{\rm yr}^{-1}$
\citep{1984ApJS...54..443P}.  Based on magnetic braking calculation by
\citet{1989ApJ...342.1019M}, the secular accretion rate for a orbital
period of 2.94 h is $\dot M = 1.1\times 10^{-9}\,\msun\,{\rm
  yr}^{-1}$.  This should be considered a lower limit, as most CVs
with periods $\sim 3-4$ h have higher secular accretion rates of $\dot
M\gtrsim 2\times 10^{-9}\,\msun\,{\rm yr}^{-1}$
\citep[e.g.,][]{Howell2001,Townsley2003}.   

Archival photographic plate photometry suggests that this star may
have been up to a magnitude brighter in the past than when imaged as
part of the Kepler Input Catalog survey, hinting that this star may
have been (or even still be) in a VY Scl-type low state.  However, the
data are sparse, and the historical magnitudes could be in error due
to the close optical companion star.

We calculate a lower limit on the distance to KIC J1924$+$4459 using the
archival 2MASS photometry, under the assumption that the $J$-band
detection is due totally to the secondary (donor) star.
\citet{Knigge2006} calculates that the donor would be of spectral type
M4.2 with $M_J=7.88$ at the edge of the period gap.  Accounting for
the Galactic extinction from \citet{1998ApJ...500..525S}, the
intrinsic distance modulus is $(m-M)_0=8.83\pm 0.14$, or
$583^{+39}_{-36}$ pc.  If, instead, we adopt the reddening of $\ebv =
0.45\pm 0.15$ derived from the Na D EW measurements, the intrinsic
distance modulus is $8.55\pm 0.19$, or $513 ^{+47}_{-43}$ pc. 

\subsection{Potential SW Sex characteristics}
The photometric properties of KIC J1924$+$4459 are consistent with,
though not exclusive to, NLs of the SW Sex class. The photometric
period is within the CV period gap, and the light curve does not show
deep eclipses.  SW Sex systems account for 55\% of CVs in the period
gap \citep{2007MNRAS.377.1747R}, and 37\% of SW Sex systems do not
show eclipses \citep{2007MNRAS.374.1359R}.  The time scale and
amplitude of the quasi-coherent variability is similar to that
observed in a number of NLs of the SW Sex class
\citep{2007MNRAS.374.1359R,2007MNRAS.377.1747R}.

One defining spectral characteristic of SW Sex stars are single-peaked
Balmer and \ion{He}{1} emission lines \citep{1995cvs..book.....W},
such as we observe in KIC J1924$+$4459.  Additional defining spectral
characteristics of the SW Sex stars include a higher-velocity emission
S-wave and variable absorption components in the trailed spectrum. We
are, unfortunately, unable to identify these features in our data due
to the combination of the low velocity amplitude, the low resolution
of these spectra, and the small phase coverage of these data.  A
longer set of time-series data at higher spectral resolution are
needed to search for this diagnostic signature.  Likewise, the phase
delay in the red-to-blue crossing of the Balmer lines that is
characteristic of many SW Sex systems cannot be tested with these data
due to the inability to determine the absolute phase of the system in
this short run of data.

In summary, we speculate that KIC J1924$+$4459 is a member of the
rapidly-growing SW Sex class of CV based on our data, which are
consistent with common definitions of SW Sex stars.  However, the data
are not sufficient to make a definitive identification at this time.

\subsection{Implications and future potential}
The discovery of a new CV, and in particular a potential SW Sex type
system, in the $2-3$ h period gap is not only important to increase
the statistics of this emerging class of accreting CVs, but also to
help understand the crucial role of angular momentum loss mechanisms
that act at these periods \citep[for details, see][and references
therein]{2009ApJ...693.1007T}.  Systems in the period gap are believed
to be transition systems. Angular momentum loss is believed to be
driven by magnetic breaking for CVs above the gap and driven by
gravitational radiation in and below the gap.  The fact that many CVs
in the gap, at least half of which are SW Sex objects, display
features indicating high $\dot M$ when this is not expected by angular
momentum loss theory is a problem that needs to be solved.  Therefore,
the study of these systems is crucial to a better understanding of CV
evolution.

This system requires further careful observations in order to 
classify the CV securely.  In particular, confirming whether KIC
J1924$+$4459 is a SW Sex-class system will require time-resolved,
high resolution spectroscopy.  (Spectro-)polarimetric and
X-ray observations can provide evidence of the presence or absence of
strong magnetic fields on the accreting white dwarf.  Long-term
photometric monitoring and a more careful analysis of archival imaging
may be able to determine if this system enters occasional VY Scl-like
low states.

The identification of a cataclysmic variable in the field of view of
the Kepler Mission provides the unprecedented opportunity to obtain
long-term, uninterrupted photometric monitoring of such a system.
First and foremost, coordinated observations combining Kepler data
with ground-based spectroscopy and/or polarimetry could help in
furthering our understanding of this system.  Additionally, Kepler's
superb photometric precision and long mission lifetime can provide a
unique opportunity to search for very low-amplitude and low-frequency
phenomena, and to open new avenues of data analysis for both this
object and cataclysmic variable stars in general.  Indeed, KIC
J1924$+$4459 is targeted for short-cadence observations by the Kepler
Asteroseismic Science Consortium (KASC) during roll 3 of the Kepler
mission; these data will soon be available. 

\acknowledgements KAW is grateful for the financial support of
National Science Foundation award AST-0602288. DDM acknowledges the
financial support of INAF PRIN-INAF-17/07.  The authors wish to thank
Arnold Klemola, Bob Hanson, Dave Monet, and Tim Brown for their great
assistance in tracking down dates of observations for the archival
photometric data.  RS wishes to thank Alfio Bonanno and Silvio Leccia,
who agreed to devote part of their TNG run to the {\it Kepler pulsator
  candidates} program following technical problems with the SARG
instrument, and the TNG technical team, Gloria Andreuzzi, Antonio
Magazzu and Luca Di Fabrizio, for the excellent job during the
service-mode observations.  Ed Sion, Paula Szkody, Steve Howell, and
James Liebert all provided insightful discussions and advice in
interpreting these results.  We are also grateful to Mukremin Kilic
and Susan Thompson for taking additional spectroscopic and photometric
data that were, alas, not used in this analysis. Fergal Mullally
kindly provided assistance in calculating barycentric corrections for
the spectroscopy. The authors thank the anonymous referee for their
time and useful comments regarding this paper.



\end{document}